\documentclass[aps,prl,twocolumn,superscriptaddress,amsmath,showpacs,
preprintnumbers]{revtex4}
\usepackage{epsfig}
\usepackage{psfrag}
\begin{document}
\preprint{\vtop{\hbox{RU06-2-B}\hbox{TKYNT-06-2}\hbox{hep-th/0603263}
\vskip24pt}}

\title{Mismatch Induced Type Transition of a Superconductor}

\author{Ioannis Giannakis}
\email{giannak@summit.rockefeller.edu}
\affiliation{Physics Department, The Rockefeller University,
1230 York Avenue, New York, NY 10021-6399}
\author{Defu Hou}
\email{hdf@iopp.ccnu.edu.cn}
\affiliation{Institute of Particle Physics, Huazhong Normal University,
Wuhan, 430079, China}
\author{Mei Huang}
\email{huang@nt.phys.s.u-tokyo.ac.jp}
\affiliation{Physics Department, University of Tokyo, Hongo, Bunkyo-ku, 
Tokyo, 113-0033}
\author{Hai-cang Ren}
\email{ren@summit.rockefeller.edu}
\affiliation{Physics Department, The Rockefeller University,
1230 York Avenue, New York, NY 10021-6399}
\affiliation{Institute of Particle Physics, Huazhong Normal University,
Wuhan, 430079, China}

\begin{abstract}
We use the Ginzburg-Landau theory near the transition temperature
in order to examine the behavior of an inhomogeneous superconductor
in the presence of a magnetic field.
We find that a transition from type I to type II superconductivity
occurs. Furthermore we calculate the critical value of the mismatch
and estimate the contributions of the fluctuations of the condensate
and the higher order terms in the Ginzburg-Landau framework.
\end{abstract}

\pacs{12.38.Mh, 24.85.+p}
\maketitle

%%%%%%%%%%%%%%%%%%%%%%%
%%% Introduction
%%%%%%%%%%%%%%%%%%%%%%%

The formation of
Cooper pairs with non-zero net momentum
is expected to occur in fermionic systems due to the presence of attractive
interactions that are responsible for the separation of the Fermi 
surfaces of the pairing electrons.
The study of inhomogeneous superconductors has been the subject
of intense theoretical and experimental research recently.
In an
electronic superconductor, the displacement of the Fermi momenta
can be caused by either the application of
a strong external magnetic field or the presence of ferromagnetic 
impurities. In cold fermionic atoms,
the mismatch of the Fermi sea occurs because the
effective masses or densities of the two species do not match.
Finally as we squeeze matter creating conditions
that exist in the interior of
compact stars, differences in the chemical potentials of
different quark flavors might lead to displacements of their
Fermi momenta since this configuration is energetically
favorable ( charge neutrality ).

The mismatch of the Fermi momenta is expected to reduce the available 
phase space for pairing. We shall denote the
difference of the kinetic energy of the pairing fermions with 
$\delta$, and we shall characterize
the strength of the pairing force in a superconductor with the 
magnitude of the energy gap $\Delta_0$, at 
zero temperature $T=0$ and zero
mismatch. The long range order will be destroyed for sufficiently high 
values of the ratio $\delta/\Delta_0$.
But prior to this a
new type of long range order is expected to emerge for 
$\delta\sim\Delta_0$. The candidate pairing states that have been 
proposed in the literature include the long sorted LOFF 
pairing state \cite{loff}, the breached
pairing state for cold atoms \cite{wilczek}, the Sarma state
for a charge neutral quark matter
\cite{huang} or a
heterogeneous separation between BCS phase and the normal phase
\cite{bedaque} for the quark matter. Figure 1
shows the possible phase diagram of an electronic superconductor with
Fermi momentum mismatch. The LOFF phase is expected to emerge within
a narrow window of the mismatch parameter beyond the line joining
$\delta_{\rm LOFF}\simeq 0.608\Delta_0$ at the transition temperature and 
$\delta_{\rm LOFF}^\prime\simeq \Delta_0/\sqrt{2}$ at $t=0$ \cite{takada}.

In this letter we shall demonstrate 
a dramatic change of the magnetic properties
of a superconductor that occurs
before the emergence of the exotic pairing states. 
More specifically, a type I 
superconductor without mismatch can evolve to a type II 
one beyond a critical value of the
ratio $\delta/\Delta_0$ when sufficiently close to the transition
temperature. This possible
crossover, sketched in the phase diagram Fig.1  by the dashed line, may
have observable consequences. Our results seem to contradict
the conventional belief
that a type II superconductor is associated to a stronger pairing force.

%\psfrag{Td1}{$T_\delta$}
%\psfrag{T0}{$T_0$}
%\psfrag{d1}{$\delta_c$}
%\psfrag{d2}{$\delta_{LOFF}$}
%\psfrag{d3}{$\delta'_{LOFF}$}
%\psfrag{d}{$\delta$}
\begin{figure}[h]
\includegraphics[scale=0.9]{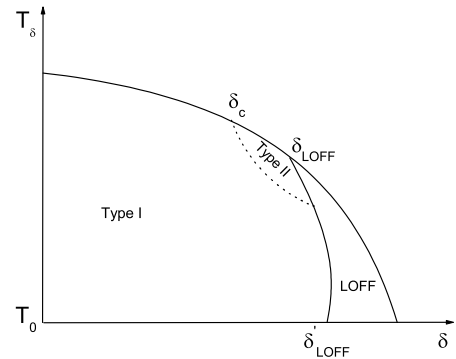}
\caption{The phase diagram on $T-\delta$ plane.}
\label{One_gluon}
\end{figure}

In the remaining of the paper, we shall derive the Ginzburg-Landau 
free energy density with a mismatch and 
we shall calculate the Ginzburg-Landau parameter 
$\kappa$. Furthermore we shall demonstrate that the
Ginzburg-Landau parameter is an increasing function of
the mismatch parameter and that it diverges
at the threshold of the LOFF state. Therefore somewhere in between, the GL 
parameter crosses the critical value $\kappa_c=1/\sqrt{2}$ that distinguishes
a type I ($\kappa<\kappa_c$) superconductor from a type II 
($\kappa>\kappa_c$).
We shall also access the robustness of this crossover within the
Ginzburg-Landau framework by estimating the fluctuation effects
and the higher order corrections. Finally, 
we shall discuss the feasibility of this proposal for realistic 
superconductors. Throughout the paper, we shall use the natural 
units with the Boltzmann constant set to one. 

We shall focus on a nonrelativistic electronic system described by the 
Hamiltonian:

\begin{equation}
H = \sum_{\vec p,s}\epsilon_{{\vec p},s}a_{\vec p,s}^\dagger a_{\vec 
p,s}-\frac{\lambda}{\Omega}
\sum_{\vec p,\vec p',\vec q}{}^\prime a_{\vec p_+\uparrow}^\dagger 
a_{-\vec p_-\downarrow}^\dagger
a_{-\vec p_-^\prime\downarrow} a_{\vec p_+^\prime\uparrow}
\label{NRHam}
\end{equation}
where $s=(\uparrow, \downarrow)$, $\epsilon_{\vec p\uparrow}
=\xi_{\vec p}+\delta$, $\epsilon_{\vec p\downarrow}=\xi_{\vec p}-\delta$, 
$\vec p_\pm=\vec p\pm\frac{\vec q}{2}$, $\vec p_\pm^\prime
=\vec p^\prime\pm\frac{\vec q}{2}$,
$\lambda$ is the strength of the interaction that is responsible for 
the pairing and $\Omega$ is the volume of the system.
Near the Fermi surface,
$\xi_p=v_F(p-k_F)$ where $k_F$ is the Fermi momentum 
and $v_F$ the Fermi velocity.
The sum $\sum_{\vec p,
\vec p^\prime, \vec q}^\prime$ of (\ref{NRHam})
extends to the pairing phase space specified
by $|\xi_p|<\omega_D$ and $|\xi_{p^\prime}|<\omega_D$ where $\omega_D$
is an intermediate energy scale between the superconducting energy gap and
the Fermi energy (Debye frequency in the case of phonon mediated pairing).
In terms of the NG representation
\begin{equation}
A_{\vec p} = \left(\begin{matrix}a_{\vec p\uparrow}\cr 
a_{-\vec p\downarrow}^\dagger\end{matrix}\right)
\end{equation}
the Hamiltonian is rewritten as
\begin{eqnarray}
H = \sum_{\vec p}\epsilon_{\vec p\downarrow}
& + & \sum_{\vec p}A_{\vec p}^\dagger
(\xi_p\sigma_3+\delta)A_{\vec p}\nonumber \\
& - & \frac{\lambda}{\Omega}\sum_{\vec p,\vec p',\vec q}{}^\prime
A_{\vec p_+}^\dagger\sigma_+A_{\vec p_-}A_{\vec p_-^\prime}
^\dagger\sigma_-A_{\vec p_+^\prime}.
\label{NRNG}
\end{eqnarray}
where $\sigma_\pm=\frac{1}{2}(\sigma_1\rm \pm i\sigma_2)$ are
the Pauli matrices.

Let's now introduce an inhomogebneous condensate, 
\begin{equation}
<A_{\vec p_-}^\dagger\sigma_-A_{\vec p_+}>=\frac{\pi^2v_F}
{\sqrt{\Omega}\lambda k_F^2\omega_D}\phi_{\vec q},
\end{equation}
and expand the 
Hamiltonian (\ref{NRNG}) to linear order of the difference
$A_{\vec p_\mp}^\dagger\sigma_\mp A_{\vec p_\pm}
-<A_{\vec p_\mp}^\dagger\sigma_\mp A_{\vec p_\pm}>$.
We recover
the mean field Hamiltonian:
\begin{eqnarray}
H_{\rm MF} & = & \sum_{\vec p}\epsilon_{\vec p\downarrow}
+\frac{1}{\lambda}\sum_{\vec q}\phi_{\vec q}^*\phi_{\vec q}
+\sum_{\vec p}A_{\vec p}^\dagger
(\xi_p\sigma_3+\delta)A_{\vec p}\nonumber \\
& - & \frac{1}{\sqrt{\Omega}}\sum_{\vec p,\vec q}^\prime
[\phi_{\vec q}^*A_{\vec p_-^\prime}^\dagger\sigma_-A_{\vec p_+^\prime}
+\phi_{\vec q}A_{\vec p_+}^\dagger\sigma_-A_{\vec p_-}].
\label{NRMF}
\end{eqnarray}
A homogeneous condensate corresponds
to $\phi_{\vec q}=\sqrt{\Omega}\Delta\delta_{\vec q,0}$
with $\Delta$ being the energy gap.

The thermodynamic potential corresponding to the
Hamiltonian (\ref{NRMF}) at temperature 
$T$ reads
\begin{equation}
{\Gamma}=\frac{1}{\lambda}\sum_{\vec q}\phi_{\vec q}^*\phi_{\vec q}
-T{\rm Tr}(\ln{\cal S}^{-1}-\ln S^{-1}),
\label{CJT}
\end{equation}
where the inverse thermal fermion propagator in the presence of 
an inhomogeneous condensate, is given by
\begin{eqnarray}
{\cal S}^{-1}(\vec p\nu|\vec p'\nu^\prime) &=& S^{-1}
(\vec p\nu|\vec p^\prime\nu^\prime)\nonumber \\
&-&\frac{1}{\sqrt{\Omega}}[\phi_{\vec p^\prime-\vec p}^*
\sigma_-+\phi_{\vec p^\prime-\vec p}\sigma_+]
\delta_{\nu\nu^\prime},
\end{eqnarray}
where 
\begin{equation}
S^{-1}(\vec p\nu|\vec p^\prime\nu^\prime)=(-i\nu+\delta+\xi_p\sigma_3)
\delta_{\vec p\vec p^\prime}\delta_{\nu\nu^\prime}
\end{equation}
is the propagator in the normal phase and $\nu$ the Matsubara energy.
The trace of (\ref{CJT}) is over the momentum, Matsubara energy and NG 
indices. The thermodynamic potential of the normal phase has been 
substracted in (\ref{CJT}).

By expanding $\Gamma$ to powers of $\phi$ and $q$ and transforming
the result to the coordinate space
\begin{equation}
\phi(\vec r)=\frac{1}{\sqrt{\Omega}}\sum_{\vec q}\phi_{\vec q}
e^{i\vec q\cdot\vec r},
\end{equation}
we obtain the
Ginzburg-Landau free energy functional with a mismatch,
\begin{equation}
\Gamma=\int d^3\vec r\Big[c|\vec\nabla\phi|^2-\alpha t|\phi|^2+\frac{1}{2}b 
|\phi|^4\Big],
\label{GL0}
\end{equation}
where
\begin{equation}
t=\frac{T_\delta-T}{T_\delta},
\end{equation} 
\begin{equation}
\alpha = \frac{k_F^2}{2\pi^2v_F}g(\delta),
\label{GLalpha}
\end{equation}
\begin{equation}
b = \frac{7\zeta(3)k_F^2}{16\pi^4v_FT_\delta^2}f(\delta),
\label{GLb}
\end{equation}
and
\begin{equation}
c=\frac{1}{6}bv_F^2.
\label{GLc}
\end{equation}
The dimensionless functions $g(\delta)$ and $f(\delta)$
are given in terms of the following expressions
\begin{equation}
g(\delta) = \Big[1+\frac{\delta}{2\pi T_\delta}{\rm Im}\psi'\Big(\frac{1}{2}
+i\frac{\delta}{2\pi T_\delta}\Big)\Big],
\end{equation}
and
\begin{equation}
f(\delta) = -\frac{1}{14\zeta(3)}{\rm Re}
\psi''\Big(\frac{1}{2}+i\frac{\delta}{2\pi T_\delta}\Big),
\end{equation}
where $\psi(z)=\Gamma^\prime(z)/\Gamma(z)$.
The transition temperature $T_\delta$ can be calculated by solving
the equation 
\begin{equation}
\ln\frac{T_\delta}{T_0}=-{\rm Re}\psi\Big(\frac{1}{2}+i\frac{\delta}{2\pi T_\delta}\Big)
-\gamma_E-2\ln 2
\end{equation}
where $T_0=\frac{e^{\gamma_E}}{\pi}\Delta_0$ is
the transition temperature at $\delta=0$.

The thermal equilibrium corresponds to the minimum of 
the Ginzburg-Landau free energy functional $\Gamma$, i.e.
$\delta\Gamma/\delta\phi(\vec r)=\delta\Gamma/\delta\phi^*(\vec r)=0$.
These conditions provide
\begin{equation}
\phi(\vec r)=\Delta=\sqrt{\frac{\alpha t}{b}}
=\sqrt{\frac{8\pi^2g(\delta)t}{7\zeta(3)f(\delta)}}T_\delta
\end{equation}
and the value of the free energy functional is
$\Gamma_{\rm min}=-\Omega\frac{\alpha^2t^2}{b}$ for sufficiently
weak mismatch.

In the absence of a mismatch, $\delta=0$, $T_\delta=T_0$, $f(0)=g(0)=1$
and the Ginzburg Landau coefficients (\ref{GLalpha}), 
(\ref{GLb}) and (\ref{GLc}) reduce to the well known expressions
\cite{gorkov}. As the mismatch $\delta$ increases, both functions
$f(\delta)$ and $g(\delta)$ decrease monotonically until 
$\delta=\delta_{\rm LOFF}$ at which point $f(\delta)$ changes sign
and the
homogeneous condensate ceases to be stable against the LOFF pairing
\cite{takada}\cite{giannakis}. We have $g(\delta)\simeq 0.140$ 
and $T_\delta\simeq 0.562T_0$ at this point.

The Ginzburg-Landau free energy of a superconductor in the presence
of a static magnetic field can be
obtained from (\ref{GL0}) by adding the magnetic energy and 
replacing 
the gradient with the covariant derivative in order to account for
the interaction of the condensate with the magnetic field. We find 
\begin{equation}
\Gamma=\int d^3\vec r\Big[\frac{1}{2}(\vec\nabla\times\vec A)^2 
+c|(\vec\nabla-2ie\vec A)\phi|^2-\alpha t|\phi|^2+\frac{1}{2}b 
|\phi|^4\Big].
\label{GL1}
\end{equation}
The magnetic properties of a superconductor are
determined by the Ginzburg-Landau 
parameter \cite{abrikosov}, 
\begin{equation}
\kappa=\frac{1}{m_A\xi}
\label{kappa}
\end{equation}
where $m_A$ is the Meissner mass and $\xi$ is the coherence length.
The Meissner mass can be calculated from (\ref{GL1}). We find that
\begin{equation}
m_A^2=\frac{7\zeta(3)e^2v_Fk_F^2\Delta^2}{12\pi^4T_\delta^2}f(\delta)
=\frac{2e^2v_Fk_F^2}{3\pi^2}g(\delta)t.
\label{meissner}
\end{equation}
In order to calculate the coherence length, we consider the perturbation 
\begin{equation}
\phi = \Delta+\frac{1}{\sqrt{2}}(u+iv)
\end{equation}
and expand $\Gamma$ to quadratic order in $u$ and $v$.
We obtain the expression
\begin{equation}
\Gamma=\Gamma_{\rm min}+\frac{1}{2}\int d^3\vec r
\Big[c(\vec\nabla u)^2+c(\vec\nabla v)^2+\frac{2}{\xi^2}u^2\Big]
\end{equation}
from which the coherence length can be identified as
\begin{equation}
\xi^2=\frac{c}{\alpha t}=\frac{7\zeta(3)v_F^2f(\delta)}
{48\pi^2g(\delta)T_\delta^2t}.
\label{coher}
\end{equation}
It follows from (\ref{kappa}) (\ref{meissner}) and (\ref{coher}) that 
the Ginzburg-Landau parameter can be written as
\begin{equation}
\kappa = \frac{1}{m_A\xi}=\frac{6\pi^2}{e\sqrt{14\zeta(3)v_F
f(\delta)}}\frac{T_\delta}{\mu}=\frac{T_\delta}{\sqrt{f(\delta)}T_0}
\kappa\mid_{\delta=0}
\end{equation}
with $\mu=\frac{1}{2}k_Fv_F$ being the Fermi energy.
While the transition temperature $T_\delta$ decreases 
slightly as $\delta$ increases,
the vanishing of $f(\delta)$ at $\delta_{\rm LOFF}$
implies that $\kappa$ diverges at that point.
Therefore for a type I superconductor at $\delta=0$, 
there exists a critical value
$\delta_c$ where $\kappa=\frac{1}{\sqrt{2}}$. The superconductor
for values of $\delta>\delta_c$
becomes type II. We find that for $\delta=\delta_c$
\begin{equation}
f(\delta_c)=\frac{9\pi^3}{7\zeta(3)\alpha_e v_F}
\Big(\frac{T_\delta}{\mu}\Big)^2.
\end{equation}
where $\alpha_e$ is the fine structure constant.

To access the robustness of this (type-I)/(type-II) crossover, 
we need to estimate the contributions of the
fluctuations of the condensate and of the higher order terms 
that were omitted
when we derived (\ref{GL0}) from (\ref{CJT}). 
According to the Ginzburg criterion, the validity of the
GL free energy (\ref{GL0}) or (\ref{GL1}) requires that
the temperature remains outside a critical window, so that the fluctuations 
pertaining to a second order phase transition can be neglected. 
One formulation
of this criterion is given in Ref.\cite{book},
more specifically
\begin{equation}
\frac{\int d^3\vec r<u(\vec r)u(0)>}{\xi^3\Delta^2}<<1
\label{GC}
\end{equation}
where the ensemble average $<u(\vec r)u(0)>$ is approximated by
\begin{equation}
<u(\vec r)u(0)>=\frac{\int[du]\exp\Big(-\frac{\Gamma}{T}\Big)u(\vec r)u(0)}
{\int[du]\exp\Big(-\frac{\Gamma}{T}\Big)}.
\end{equation}
We find that
\begin{equation}
<u(\vec r)u(0)>=\frac{T}{4\pi c r}e^{-\frac{\sqrt{2}r}{\xi}} 
\end{equation}
and (\ref{GC}) becomes
\begin{equation}
\frac{T_c}{c\xi\Delta^2}<<1.
\end{equation}
Substituting the expressions for $c$, $\xi$ and $\Delta$, we obtain 
the following criterion for ignoring the fluctuations of the condensate
\begin{equation}
t>>\frac{864\pi^6}{7\zeta(3)f(\delta)g(\delta)}\frac{T_c^4}{\mu^4}.
\label{critical}
\end{equation}
Notice that, as the mismatch increases
the width of the critical window diverges at $\delta_{\rm LOFF}$,
indicating that the fluctuations of the condensate cannot
be ignored.
This is in aggreement with the conjecture of reference \cite{mei}.

Let's now assume approximately equal contributions 
for the three 
subleading terms, $q^4$, $q^2\Delta^4$ and $\Delta^6$, and
let's choose the correction of the $\Delta^6$ 
as the representative. A detailed calculation shows that 
the ratio of this term with the contribution
from the quartic term of (\ref{GL0}) is
\begin{equation}
\frac{1}{12}\Big(\frac{\Delta}{2\pi T_\delta}\Big)^2
\frac{{\rm Re}\psi^{''''}\Big(\frac{1}{2}+i\frac{\delta}{2\pi T_\delta}\Big)}
{{\rm Re}\psi^{''}\Big(\frac{1}{2}+i\frac{\delta}{2\pi T_\delta}\Big)}.
\end{equation}
The condition for it to be small implies that
\begin{equation}
t<<\frac{49\zeta^2(3)}{62\zeta(5)}\frac{f^2(\delta)}{g(\delta)},
\label{corr}
\end{equation}
where we have replaced $-{\rm Re}\psi^{''''}(...)$ with its maxmum value at 
$\delta=0$, $744\zeta(5)$. 
By combininig (\ref{critical}) and (\ref{corr}), we arrive
at the conclusion that the validity of the GL approach 
at the crossover is justified if
\begin{equation}
\alpha_ev_F<<0.71\Big(\frac{T_\delta}{\mu}\Big)^{\frac{2}{3}}.
\label{margin}
\end{equation}
For Al, we have $T_0=1.18$K, $\mu=1.35\times 10^5$K
and $v_F=2.02\times 10^8$cm/s
\cite{kittel}, the l.h.s. of (\ref{margin}) is about $4.9\times 10^{-5}$ 
while the r. h. s. is about $2.0\times 10^{-4}$. In the case of Pb,
we have $T_0=7.19$K,
$\mu=1.09\times 10^5$K and $v_F=1.82\times 10^8$cm/s\cite{kittel},
the l.h.s. is about
$4.4\times 10^{-5}$ while the r. h. s. is about $7.9\times 10^{-4}$.
In view of numerical uncertainties behind the formulation of the 
Ginzburg criterion (\ref{GC}), it is plausible that the (type-I)/(type-II) 
crossover can be marginally described by the GL of (\ref{GL1}). 

The situation at $T=0$ is completely different. The Meissner mass at 
$T=0$ but $\delta\neq 0$ was calculated in the
literature \cite{shovkovy} and was
found to remain constant for $\delta<\Delta$. There are two coherence 
lengths: the Pippard length, which measures the size of the Cooper pair 
and the length associated with
the Anderson-Higgs mass that corresponds to the susceptibility of the 
free energy to an inhomogeneous variation of the gap magnitude. Both 
of them are of order $\xi\sim\frac{v_F}{\Delta}$ at $\delta=0$.
For $\delta\neq 0$, our preliminary results \cite{ioannis} indicate that 
both of them remain constant for $\delta<\Delta$, in a similar
manner with the Meissner
mass. Therefore, the crossover between type I and type II 
superconductors does
not extend to $T=0$. A type I superconductor remains type I until 
$\delta\simeq\frac{1}{\sqrt{2}}\Delta_0$ at which
point LOFF pairing becomes energetically favorable.

The critical magnetic field for a type I superconductor is given by 
\begin{equation}
H_c^2=\frac{\alpha^2t^2}{b}=\frac{g^2(\delta)}{f(\delta)}
H_c^2\mid_{\delta=0}.
\label{hc}
\end{equation}
The magnitude of $H_c$ at $T=0$ and $\delta=0$ is below 1000G for most 
type I superconductors. The magnetic field necessary
for Zeeman splitting
with $\delta\simeq\delta_{\rm LOFF}$ is approximately
$\frac{1.76\times 10^4}{g}$G, where
$g$ is the gyromagnetic ratio of the electrons in 
the metal. Even with the enhanced value of the
critical field given by (\ref{hc}) for $\delta$ towards $\delta_{\rm LOFF}$, 
it is unlikely that the (type-I)/(type-II) crossover can be implemented with 
a strong external magnetic field without destroying superconductivity.
On the other hand, the exchange field of ferromagnetic impurities may achieve
this type of transition. It's action on 
the electron spins is equivalent to that of a magnetic field 
capable of causing a significant mismatch, but at the same time 
remaining well below $H_c$.
The magnetic field by asymmetric electron spins can hardly exceeds few G, 
which can be neglected.

In this letter, we developed the Ginzburg-Landau theory of an electronic 
superconductor at weak couling when the Fermi momenta of opposite 
spins are mismatched.
A remarkable feature that we found is that
a type I superconductor becomes type II
at sufficiently high values of the mismatch
near the transition temperature. We have also estimated the 
fluctuation effects and the higher order corrections 
to the GL framework and argued that this 
type of transition is likely to be robust.

M. H. would like to thank T. Hatsuda and T. Matsuura for helpful discussions
on coherent lengths. 
The work of I.G. and H.C.R is supported in part by US Department of Energy 
under grants DE-FG02-91ER40651-TASKB. The work of D.F.H. is supported in part 
by NSFC under grant No.10135030 and Educational Committee under grant No. 
704035. the work of M.H. is supported in part by the Japan Society for the 
Promotion of Science Fellowship Program. The work of D.F.H and H.C.R. is also
supported in part by NSFC under grant No. 10575043.

\end{document}